\RequirePackage{fix-cm}
\documentclass[smallextended]{svjour3}       % onecolumn (second format)
\usepackage{mathptmx}      % use Times fonts if available on your TeX system
\usepackage{graphicx}
\usepackage{braket}
\usepackage{cite}
\usepackage[utf8]{inputenc}
\usepackage{amssymb}
\usepackage{amsmath}
\usepackage{hyperref}
\usepackage{mathdots}
\usepackage{qcircuit}
\usepackage{etoolbox}
\usepackage{placeins}
\usepackage[left=3.4cm,right=3.4cm,bottom=3.5cm,top=3.4cm]{geometry}

\usepackage[titletoc,title]{appendix}
\usepackage{etoolbox}
\apptocmd{\appendices}{\apptocmd{\thesection}{: }{}{}}{}{}

\newcommand{\ggate}[1]{*+<1.4em>{\phantom{#1}} \POS ="i","i"+R; **\dir{-}; \qw}
\newcommand{\argmax}{\mathop{\rm arg~max}\limits}

\journalname{Quantum Information Processing}

\def\makeheadbox{{%
\hbox to0pt{\vbox{\baselineskip=10dd\hrule\hbox
to\hsize{\vrule\kern3pt\vbox{\kern3pt
\hbox{This is a post-peer-review, pre-copyedit version of an article published in QINP.}
\hbox{The final authenticated version is available online at: \href{[Insert doi here]}{https://doi.org/10.1007/s11128-019-2565-2}.}
\kern3pt}\hfil\kern3pt\vrule}\hrule}%
\hss}}}

\begin{document}

\title{Amplitude Estimation without Phase Estimation%\thanks{Grants or other notes
\thanks{Y. Suzuki, S. Uno: Equally contributing authors.}
}
%\authorrunning{}

\author{Yohichi Suzuki \and
        Shumpei Uno \and
        Rudy Raymond\and \\
        Tomoki Tanaka \and
        Tamiya Onodera\and
        Naoki Yamamoto
        }
\institute{Yohichi Suzuki \and Shumpei Uno \and Rudy Raymond \and Tomoki Tanaka \and Tamiya Onodera \and Naoki Yamamoto
	\at Quantum Computing Center, Keio University, 3-14-1 Hiyoshi, Kohoku-ku, Yokohama, Kanagawa, 223-8522, Japan
           \and
          Shumpei Uno
           \at Mizuho Information \& Research Institute, Inc., 2-3 Kanda-Nishikicho, Chiyoda-ku, Tokyo, 101-8443, Japan
           \and
           Rudy Raymond \and Tamiya Onodera
            \at IBM Research - Tokyo, 19-21 Nihonbashi Hakozaki-cho, Chuo-ku, Tokyo, 103-8510, Japan
            \and
            Tomoki Tanaka
             \at   Mitsubishi UFJ Financial Group, Inc. and MUFG Bank, Ltd., 2-7-1 Marunouchi, Chiyoda-ku, Tokyo, 100-8388, Japan
            \and
             Naoki Yamamoto
             \at Department of Applied Physics and Physico-Informatics, Keio University, 3-14-1 Hiyoshi, Kohoku-ku, Yokohama, Kanagawa, 223- 8522, Japan
             \email{yamamoto@appi.keio.ac.jp} 
                 }

\date{Received: date / Accepted: date}


\maketitle

%\footnotetext [1]{cairo university \cite{bla}.}

\begin{abstract}
This paper focuses on the quantum amplitude estimation algorithm, which is a core subroutine in quantum computation for various applications. The conventional approach for amplitude estimation is to use the phase estimation algorithm, which consists of many controlled amplification operations followed by a quantum Fourier transform. However, the whole procedure is hard to implement with current and near-term quantum computers. In this paper, we propose a quantum amplitude estimation algorithm without the use of expensive controlled operations; the key idea is to utilize the maximum likelihood estimation based on the combined measurement data produced from quantum circuits with different numbers of amplitude amplification operations. Numerical simulations we conducted demonstrate that our algorithm asymptotically achieves nearly the optimal quantum speedup with a reasonable circuit length. 
\end{abstract}

\section{Introduction}
Quantum computers are expected to allow us to perform high-speed computations over classical computations for problems in a wide range of scientific and technological fields. Environments in which quantum algorithms can be executed by real quantum devices are currently being provided \cite{IBMQ2019,PhysRevX.8.021012,PhysRevLett.119.180511}. Real quantum devices with several tens of qubits will soon be realized in near future, although those are the so-called noisy intermediate-scale quantum (NISQ) devices that impose several practical limitations on their use \cite{Preskill2018}, both in the number of gate operations and the number of available qubits. Hence, several custom subroutines taking into account these constraints have been proposed, typically the variational quantum eigensolver \cite{McClean2016,Yung2014}.\par

In this paper, we focus on the amplitude estimation algorithm, which is a core subroutine in quantum computation for various applications, e.g., in chemistry \cite{Knill2007,Guzik2008}, finance \cite{Bromley2018,Woerner2019}, and machine learning \cite{Wiebe2015,Wiebe2016a,Wiebe2016b,Kerenidis2018}. In particular, quantum speedup of Monte Carlo sampling via amplitude estimation \cite{Montanaro2015} lies in the heart of these applications. Therefore, in light of its importance, we followed the aforementioned direction and developed a new amplitude estimation algorithm that can be executed in NISQ devices. \par

Note that Ref.~\cite{Brassard2002} demonstrated that the amplitude estimation problem can be formulated as a phase estimation problem \cite{kitaev1996quantum}, where the amplitude to be estimated is inferred from the eigenvalue of the corresponding amplification operator. Owing to the ubiquitous nature of the eigenvalue estimation problem, some versions of the phase estimation algorithm suitable for NISQ devices \cite{Svore2013,Granade2016,OBrien2018,denBerg2019,CRWie2019} have been proposed (with the last one appeared slightly after ours), and they all rely on classical post-processing statistics such as the Bayes method. However, these modified phase estimation algorithms as well as the original scheme \cite{kitaev1996quantum} still involve many controlled operations (e.g., the controlled amplification operation in the case of Ref.~\cite{Brassard2002}) that can be difficult to implement on NISQ devices. Therefore, a new algorithm specialized to the amplitude estimation problem is required, one that does not use expensive controlled operations.\par

The goal of amplitude estimation is, in its simplest form, to estimate the unknown parameter $\theta$ contained in the state $\left|\Psi\right> = \sin\theta \left|{\rm good}\right> + \cos\theta \left|{\rm bad}\right>$, where $\left|{\rm good}\right>$ and $\left|{\rm bad}\right>$ are given orthogonal state vectors. Our scheme is composed of the amplitude amplification process and the maximum likelihood (ML) estimation; the controlled operations and the subsequent quantum Fourier transform (QFT) are not involved. The amplification process transforms the coefficient of $\left|{\rm good}\right>$ to $\sin((2m+1) \theta)$ with $m$ being the number of operations; if $\theta$ is known, then, by suitably choosing $m$, we can enhance the probability of hitting ``good'' quadratically greater than the classical case, where no amplitude amplification is utilized \cite{Grover1996}. However, the function $\sin((2m+1) \theta)$ does not always take a relatively large value for a certain $m$ because $\theta$ is unknown in this problem, meaning that an effective quantum speedup is not always available; also, the ML estimate is not uniquely determined due to the periodicity of this function. Our strategy is the first to make measurements on the transformed quantum state and to construct likelihood functions for several $m$, say $\{m_0, \ldots, m_M\}$, and then combine them to construct a single likelihood function that uniquely produces the ML estimate; see Fig.~\ref{amplitude_estimation_diagram}. The broad concept behind this scheme is to combine the data produced from different quantum circuits, and the scheme might be performed even on NISQ devices to compute a target value faster than classical algorithms via some post-processing. Actually a numerical simulation demonstrates that, by appropriately designing $\{m_k\}$, compared with the classical sampling we can achieve nearly a square-root reduction in the total number of queries to reach the specified estimation precision; notably, only relatively short-depth circuits are required to achieve this quantum speedup. We also show that, in an application of the amplitude estimation to Monte Carlo integration, our algorithm requires many fewer controlled NOT (CNOT) gates than the conventional phase-estimation-based approach, so it is suitable for obtaining quantum advantages with NISQ devices. Note that Ref.~\cite{Abrams1999} also took the approach without using the phase estimation method, but it needed to change the query in each iteration, which is highly demanding in practice. Also the paper Ref.~\cite{Zintchenko2016} gave an amplitude estimation scheme that employs a Bayes rule together with applying random Unitary operations (subjected to the Haar measure) to ideally realize the quadratic speedup, without a controlled Unitary operation; this scheme is applicable to low-dimensional quantum circuits, due to the hardness to implement the random Unitaries.

\section{Preliminary\label{sec:preliminary}}
We herein briefly describe the quantum amplitude amplification, which is the basis of our approach for the amplitude estimation problem.\par

Our proposed algorithm mainly consists of two parts: quantum amplitude amplification and amplitude estimation based on likelihood analysis. The amplitude amplification \cite{Brassard1997,Grover1998} is the generalization of the Grover's quantum searching algorithm \cite{Grover1996}. Similar to quantum searching, the amplitude amplification is known to achieve quadratic speedup over the corresponding classical algorithm.\par

We assume a unitary operator $\mathcal{A}$ that acts on $(n+1)$ qubits, such that $\ket{\Psi} = \mathcal{A} \ket{0}_{n+1} = \sqrt{a}\ket{\tilde{\Psi}_1}\ket{1}+\sqrt{1-a}\ket{\tilde{\Psi}_0}\ket{0}$, where $a \in \left[0,1\right]$ is the unknown parameter to be estimated, while $\ket{\tilde{\Psi}_1}$ and $\ket{\tilde{\Psi}_0}$ are the $n$-qubit normalized good and bad states. The query complexity of estimating $a$ is counted by the number of the operations of $\mathcal{A}$, which is often denoted as the number of {\it{queries}} for simplicity. By performing measurements on $\ket{\Psi}$ repeatedly, we can infer $a$ from the ratio of obtaining the good and bad states, but the number of queries is exactly the same as the classical one in this case.\par

The advantage offered by the quantum amplitude amplification is that, instead of measuring right after the single operation of $\mathcal{A}$, we can amplify the probability of obtaining the good state by applying the following operator.
\begin{equation}
\mathbf{Q}=-\mathcal{A}\mathbf{S}_0\mathcal{A}^{-1}\mathbf{S}_{\chi},\label{Q_def}
\end{equation}
where the operator $\mathbf{S}_\chi$ puts a negative sign to the good state, i.e., $\mathbf{S}_\chi\ket{\tilde{\Psi}_1}\ket{1} = - \ket{\tilde{\Psi}_1}\ket{1}$, and does nothing to the bad state. Similarly, $\mathbf{S}_0$ puts a negative sign to the all-zero state $\ket{0}_{n+1}$ and does nothing to the other states. $\mathcal{A}^{-1}$ is the inverse of $\mathcal{A}$, the operation of which requires the same query complexity as $\mathcal{A}$.\par

By defining a parameter $\theta_a \in \left[0,\pi/2\right]$ such that $\sin^2{\theta_a} = a$, we have
\begin{equation}
\mathcal{A} \ket{0}_{n+1} = \sin{\theta_a}\ket{\tilde{\Psi}_1}\ket{1}+\cos{\theta_a}\ket{\tilde{\Psi}_0}\ket{0}.\label{eq:psi}
\end{equation}
Brassard et al. \cite{Brassard2002} showed that repeatedly applying $\mathbf{Q}$ for $m$ times on $\ket{\Psi}$ results in
\begin{equation}
\mathbf{Q}^m\ket{\Psi} = \sin((2m+1)\theta_a)\ket{\tilde{\Psi}_1}\ket{1}+\cos((2 m+1)\theta_a)\ket{\tilde{\Psi}_0}\ket{0}.\label{Qj2}
\end{equation}
This equation represents that, after applying $\mathbf{Q}$ $m$ times (with $2m$ queries), we can obtain the good state with a probability of at least $4m^2$ times larger than that obtained from $\mathcal{A}\ket{0}_{n+1}$ for sufficiently small $a$. This is in contrast with having $2m$ number of measurements from $\mathcal{A}\ket{0}_{n+1}$, which only gives the good state with probability $2m$ times larger. This intuitively gives the quadratic speedup obtained from the amplitude amplification: if we can infer the ratio of the good state after the amplitude amplification, we can estimate the value of $a$ from the number of queries required to obtain such a ratio.\par

The conventional amplitude estimation\cite{Brassard2002} utilizes the quantum phase estimation which requires a quantum circuit that implements the multiple controlled $\mathbf{Q}$ operations, namely, $\mbox{Controlled-}\mathbf{Q}:~\ket{m}\ket{\Psi} \rightarrow \ket{m}\mathbf{Q}^m\ket{\Psi}$. Performing the controlled operations simultaneously on many $m$'s consecutively and gathering the amplitude by the inverse QFT enables an accurate estimation of $a$ \cite{Brassard2002}. However, this approach suffers from the need for many controlled gates (thus, CNOT gates) and additional ancilla qubits (the number of which is dictated by the required accuracy). Such an approach can be problematic for NISQ devices.

\section{Amplitude estimation without phase estimation}
\subsection{Algorithm}
\begin{figure}[tb]
\centering
\includegraphics[scale=1]{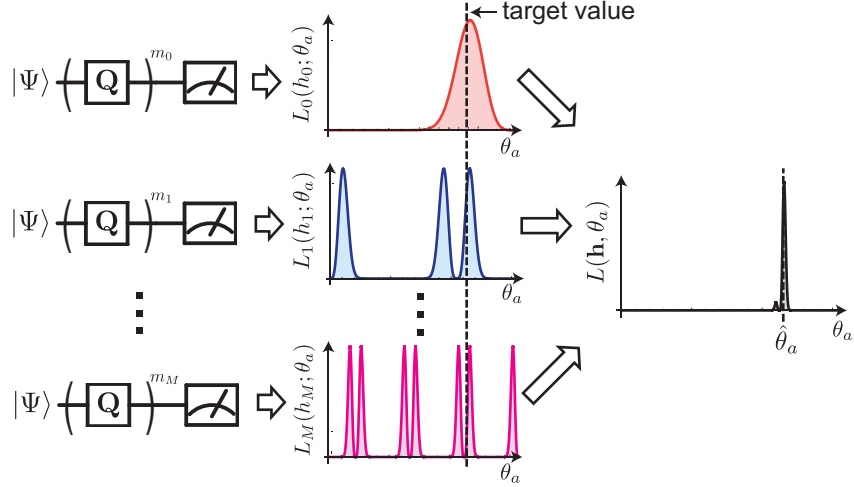}
\caption{Schematic picture of our amplitude estimation algorithm using the ML estimation. After preparing the states $\mathbf{Q}^{m_k} \ket{\Psi}$, the numbers of measuring good states, i.e. $h_k$ are obtained (left). Based on the obtained $h_k$, the likelihood functions $L_k(h_k;\theta_a)$ are constructed (center). Finally, a single likelihood function $L(\mathbf{h};\theta_a)$ is introduced by combining the likelihood functions $L_k(h_k;\theta_a)$ (right). The ML estimate is the value that maximizes the likelihood function $L(\mathbf{h};\theta_a)$.}
\label{amplitude_estimation_diagram}
\end{figure}

This section shows the quantum algorithm to estimate $\theta_a$ in Eq.~(\ref{Qj2}) without using the conventional phase-estimation-based method \cite{Brassard2002}. The first stage of the algorithm is to make good or bad measurements on the quantum state ${\mathbf{Q}}^{m_k}\ket{\Psi}$ for a chosen set of $\{m_k\}$. Let $N_k$ be the number of measurements (shots) and $h_k$ be the number of measuring good states for the state ${\mathbf{Q}}^{m_k}\ket{\Psi}$; then, because the probability measuring the good state is $\sin^2((2m_k+1)\theta_a)$, the likelihood function representing this probabilistic event is given by 

\begin{equation}
\label{LHpart}
 L_k(h_k;\theta_a) 
 = \left[\sin^2((2 m_k+1)\theta_a)\right]^{h_k}
 \left[\cos^2((2 m_k+1)\theta_a)\right]^{N_k-h_k}. 
\end{equation}
The second stage of the algorithm is to combine the likelihood functions $L_k(h_k;\theta_a)$ for several $\{m_0, \ldots, m_M\}$ to construct a single likelihood function $L(\vec{h};\theta_a)$: 

\begin{equation}
\label{LH}
 L(\vec{h};\theta_a) = \prod_{k=0}^{M}L_k(h_k;\theta_a), 
\end{equation}
where $\vec{h}=(h_0,h_1,\cdots,h_M)$. The ML estimate is defined as the value that maximizes $L(\vec{h};\theta_a)$: 

\begin{equation}
\label{ML estimate}
 \hat{\theta}_a 
 = \argmax_{\theta_a} ~ L(\vec{h};\theta_a)
 = \argmax_{\theta_a} ~ \ln L(\vec{h};\theta_a) 
\end{equation}
The whole procedure is summarized in Fig.~\ref{amplitude_estimation_diagram}. Now $a$ and $\theta_a$ are uniquely related through $a=\sin^2\theta_a$ in the range $0\leq\theta_a\leq\pi/2$, and $\hat{a}:=\sin^2\hat{\theta}_a$ is the ML estimate for $a$; thus, in what follows, $L(\vec{h}; a)$ is denoted as $L(\vec{h}; \theta_a)$. Note that the random variables $h_0, h_1, \ldots, h_M$ are independent but not identically distributed because the probability distribution for obtaining $h_k$, i.e., $p_k(h_k;\theta_a)\propto L_k(h_k;\theta_a)$, is different for each $k$; however, the set of multidimensional random variables $\vec{h}=(h_0,h_1,\cdots,h_M)$ is independently generated from the identical joint probability distribution $p(\vec{h};\theta_a)\propto L(\vec{h};\theta_a)$.

This algorithm has two caveats: (i) if only a single amplitude amplification circuit is used like in the Grover search algorithm, i.e., the case $M=0$ and $m_0\ne0$, the ML estimate $\hat{\theta}_a$ cannot be uniquely determined due to the periodicity of $L_0(h_0;\theta_a)$, and (ii) if no amplification operator is applied, i.e., $m_k=0~\forall k$, then the ML estimate is unique, but it does not have any quantum advantages, as shown later. Hence, the heart of our algorithm can be regarded as the {\it quantum circuit fusion} technique that combines some quantum circuits to determine the target value uniquely, while some quantum advantage is guaranteed. 

\subsection{Statistics: Cram\'er--Rao bound and Fisher information}
The remaining to be determined in our algorithm was to design the sequences $\{m_k, N_k\}$ so that the resulting ML estimate $\hat{\theta}_a$ might have a distinct quantum advantage over the classical one. Here, we introduce a basic statistical method to carry out this task and, based on that method, give some specific choice of $\{m_k, N_k\}$. \par

First, in general, the {\it Fisher information} $\mathcal{I}(a)$ is defined as 

\begin{eqnarray}
\label{Fisher}
	\mathcal{I}(a) 
 = {\mathbb{E}}\left[ \left( 
 \frac{\partial}{\partial a}\ln L(x;a) \right)^2 \right],
\end{eqnarray}
where the expectation is taken over a random variable $x$ subjected to a given probability distribution $p(x;a)$ with an unknown parameter $a$. The importance of the Fisher information can be clearly seen from the fact that any estimate $\hat a$ satisfies the following {\it Cram\'er--Rao inequality}.

\begin{eqnarray}
\label{CR}
{\rm{var}}(\hat{a}) = \mathbb{E}[(\hat{a}-\mathbb{E}[\hat{a}])^2]
	 \geq \frac{[1+b'(a)]^2}{{\mathcal{I}}(a)},
\end{eqnarray}
where $b(a)$ represents the bias defined by $b(a)=\mathbb{E}[\hat{a}-a]$ and  $b'(a)$ indicates the derivative of $b(a)$ with respect to $a$. It is easy to see that the mean squared estimation error satisfies 

\begin{eqnarray}
\label{MSE}
 \mathbb{E}[(\hat{a}-a)^2]\geq\frac{[1+b'(a)]^2}{{\mathcal{I}}(a)}+b(a)^2.
\end{eqnarray}
A specifically important property of the ML estimate, which maximizes the likelihood function $\prod_{k}p({x}_k;a)$ with the measurement data ${x}_k$, is that it becomes unbiased, i.e., $b(a)=0$, and further achieves the equality in Eq.~\eqref{MSE} in the large number limit of measurement data \cite{rao1973linear}; that is, the ML estimate is asymptotically optimal. \par

In our case, by substituting Eqs.~\eqref{LHpart} and \eqref{LH} into Eq.~\eqref{Fisher} together with a straight forward calculation $\mathbb{E}[h_k]=N_k\sin^2((2m_k+1)\theta_a)$, we find that 

\begin{equation}
\label{Fisher_final}
 {\mathcal{I}(a)}=\frac{1}{a(1-a)}\sum_{k=0}^{M}{N_k}(2 m_k+1)^2.
\end{equation}
Also, for any sequences $\{m_k, N_k\}$, the total number of queries is given as

\begin{equation}
\label{calling}
 N_{\rm{q}}=\sum_{k=0}^M N_k (2m_k+1).
\end{equation}
As stated before, the coefficient $2$ multiplying $m_k$ in Eq.~(\ref{calling}) originates from the fact that the operator $\mathbf{Q}$ uses $\mathcal{A}$ and $\mathcal{A}^{-1}$, and the constant $+1$ is due to the initial state preparation of $\ket{\Psi}=\mathcal{A}\ket{0}_{n+1}$. If $\mathbf{Q}$ is not applied to $\ket{\Psi}$ and if only the final measurements are performed for $\ket{\Psi}$, i.e., $m_k=0$ for all $k$, the total number of queries is identical to that of classical random sampling. Because $N_k$ and $(2 m_k + 1)$ are positive integers, the Fisher information in Eq.~\eqref{Fisher_final} satisfies the following relation.

\begin{equation}
\label{Fisher_ineq}
 {\mathcal{I}}(a) 
 ~\le~ \frac{1}{a(1-a)}\left(\sum_{k=0}^M N_k(2m_k+1)\right)^2 
 ~=~ \frac{1}{a(1-a)}N_{\rm{q}}^2.
\end{equation}
Here, $\hat a$ is set to the ML estimate \eqref{ML estimate}, and the estimation error is considered to be $\hat{\epsilon}=\sqrt{\mathbb{E}[(\hat{a}-a)^2]}$ in this case. The total number of measurements $\sum_{k=0}^M N_k$ is assumed to be sufficiently large, in which case the ML estimate asymptotically converges to an unbiased estimate and achieves the lower bound of the Cram\'er--Rao inequality \eqref{CR}, as aforementioned. Hence, from Eqs.~\eqref{CR} and \eqref{Fisher_ineq}, the error $\hat{\epsilon}$ satisfies

\begin{eqnarray}
\label{lowest}
 \hat{\epsilon} 
 \rightarrow \frac{1}{{\mathcal{I}}(a)^{1/2}}
 \ge \frac{\sqrt{a(1-a)}}{N_{\rm{q}}}. 
\end{eqnarray}
(More precisely, $\hat{\epsilon} \, \mathcal{I}(a)^{1/2} \rightarrow 1$.) That is, the lower bound of the estimation error is on the order of $\mathcal{O}(N_{\rm{q}}^{-1})$, which is referred to as the Heisenberg limit. This is in stark contrast to the classical sampling method, the estimation error of which is lower bounded by $\sqrt{a(1-a)}/N_{\rm{q}}^{1/2}$, obtained by setting $m_k=0~\forall k$ (i.e., a case with no amplitude amplification) in Eqs.~\eqref{Fisher_final} and \eqref{calling}; that is, the lower bound is at best on the order of $\mathcal{O}(N_{\rm{q}}^{-1/2})$ in the classical case. \par

Now, we can consider the problem posed at the beginning of this subsection: designing the sequences $\{m_k, N_k \}$ so that the resulting ML estimate $\hat{\theta}_a$ outperforms the classical limit $\mathcal{O}(N_{\rm{q}}^{-1/2})$ and hopefully achieves the Heisenberg limit $\mathcal{O}(N_{\rm{q}}^{-1})$, i.e., the quantum quadratic speedup. Although the problem can be formulated as a maximization problem of Fisher information \eqref{Fisher_final} with respect to $\{ m_k, N_k \}$ under some constraints on these variables, here we fix $N_k$'s to a constant and provide just two examples of the sequence $\{ m_k\}$: 

\begin{itemize}
  \item{Linearly incremental sequence (LIS): $N_k=N_{\rm{shot}}$ for all $k$, and $m_k=k$, i.e., it increases as $m_0=0, m_1=1, m_2=2, \cdots, m_M=M$}. 
  \item{Exponentially incremental sequence (EIS): $N_k=N_{\rm{shot}}$ for all $k$, and $m_k$ increases as $m_0=0, m_1=2^0, m_2=2^1,\cdots, m_M=2^{(M-1)}$}. 
\end{itemize}
In the case of LIS, the Fisher information \eqref{Fisher} and the number of queries \eqref{calling} are calculated as ${\mathcal{I}}(a)=N_{\rm{shot}}(2M+3)(2M+1)(M+1)/(3a(1-a))$ and $N_{\rm{q}}=N_{\rm{shot}}(M+1)^2$, respectively. Because $N_{\rm{q}}\sim N_{\rm{shot}}M^2$ and ${\mathcal{I}}(a)\sim N_{\rm{shot}}M^3/(3 a(1-a))$ when $M\gg 1$, the lower bound of the estimation error is evaluated as $\hat{\epsilon}=1/{\mathcal{I}}(a)^{1/2}\sim N_{\rm{q}}^{-3/4}$; hence, a distinct quantum advantage occurs, although it does not reach the Heisenberg limit. Next for the case of EIS, we find $N_{\rm{q}}\sim N_{\rm{shot}} 2^{M+1}$ and $\mathcal{I}(a)\sim N_{\rm{shot}} 2^{2(M+1)}/3$, which as a result lead to $\hat{\epsilon}\sim N_{\rm{q}}^{-1}$. Therefore, this choice is asymptotically optimal; we again emphasize that the statistical method certainly serves as a guide for us to find an optimal sequence $\{m_k\}$, achieving an optimal quantum amplitude estimation algorithm. But note that these quantum advantages are guaranteed only in the asymptotic regime and that the realistic performance with the finite (or rather short) circuit depth should be analyzed. We will carry out a numerical simulation to see this realistic case in the following. 

\subsection{Numerical simulation}
In this section, the ML estimates $\hat{\theta}_a$ and errors $\hat{\epsilon}$ are evaluated numerically for several fixed target probabilities $a=\sin^2 \theta_a$. Based on the chosen sequence of $\{N_k\}$ and $\{m_k\}$ shown in the previous subsection, $h_k$'s in Eq.~\eqref{LH} are generated using the Bernoulli sampling with probability $\sin^2((2m_k+1)\theta_a)$ for each $k$. The global maximum of the likelihood function can be obtained by using a modified brute-force search algorithm; the global maximum of $\prod_{k=0}^{m}L_k(h_k;\theta_a)$ is determined by searching around the vicinity of the estimated global maximum for $\prod_{k=0}^{m-1}L_k(h_k;\theta_a)$. The errors $\hat{\epsilon}$ are evaluated by repeating the aforementioned procedures 1000 times for each $N_{\rm{q}}$.\par

\begin{figure}[htb]
\centering
\includegraphics[scale=1]{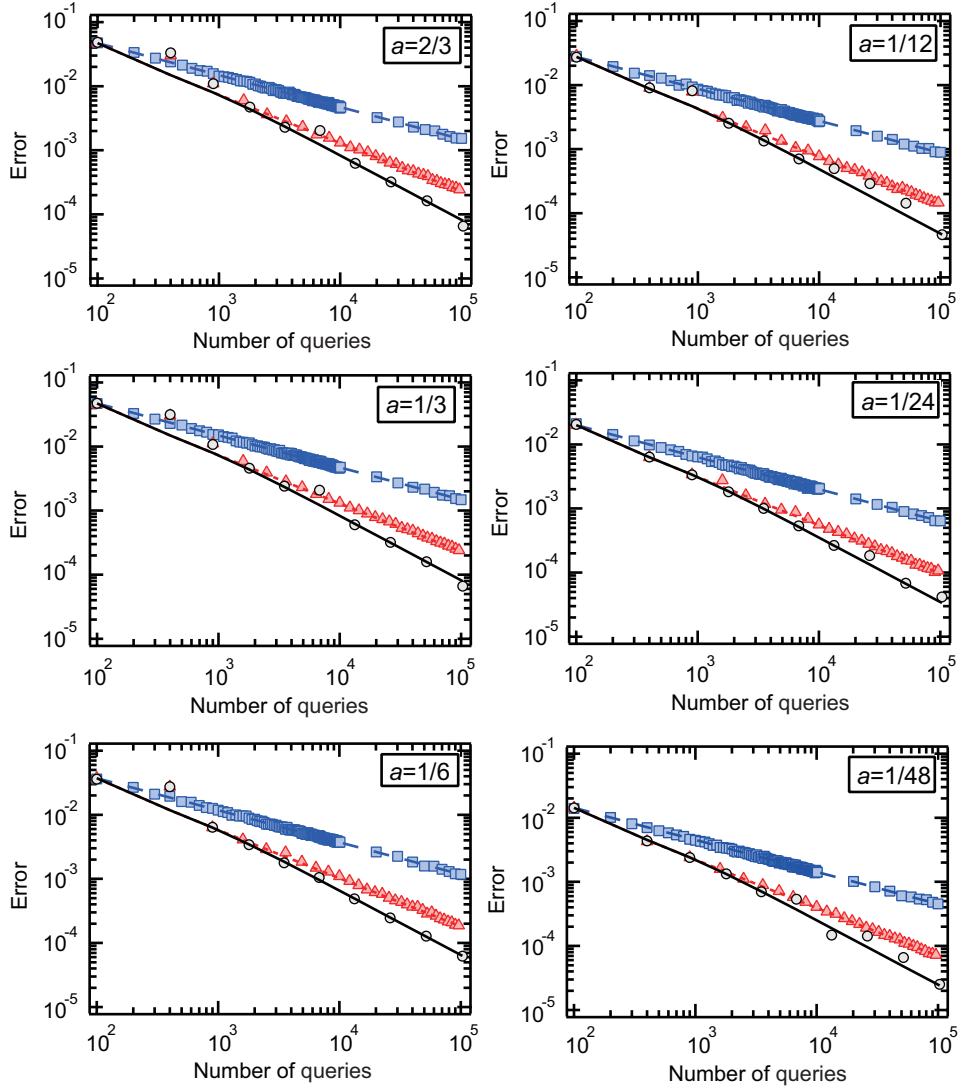}
\caption{Relationships between the number of queries and the estimation error for several target probabilities $a=\sin^2\theta_a$. The lower bounds of estimation error based on the Cram\'{e}r--Rao inequality are depicted as lines, the (blue) dashed line is for $m_k=0$ for all $k$ (classical random sampling), the (red) dotted line is for  $m_0=0, m_1=1, \cdots,  m_M=M$ (LIS), and the (black) solid line is for $m_0=0, m_1=2^0, \cdots, m_M=2^{M-1}$ (EIS), respectively. The estimation errors obtained by numerical simulations are also plotted as symbols, the (blue) squares are for classical random sampling, the (red) triangles are for LIS, and the (black) circles are for EIS.}
\label{num_vs_orac_vs_error}
\end{figure}

In Fig.~\ref{num_vs_orac_vs_error}, the relationship between the number of queries and errors are plotted for the target probabilities $a=\sin^2 \theta_a=2/3$, $1/3$, $1/6$, $1/12$, $1/24$, and $1/48$ with $N_{\textrm{shot}}=100$. The (red) triangles and (black) circles in Fig.~\ref{num_vs_orac_vs_error} are errors that are obtained using LIS and EIS, respectively. For comparison, numerical simulations with $m_k=0$ for all $k$ are also performed, and the results are plotted as (blue) squares in Fig.~\ref{num_vs_orac_vs_error}. In addition, the lower bounds of errors \eqref{lowest} when the estimate is not biased are also plotted as (red) dotted and (black) solid lines for LIS and EIS, respectively. The (blue) dashed lines in Fig.~\ref{num_vs_orac_vs_error} are the lower bounds for classical random sampling, i.e., $\sqrt{a(1-a)/N_{\rm{q}}}$.\par

The slopes of the simulated results with the target probability $a=\sin^2\theta_a=1/48$ ranging from $N_{\rm{q}}\simeq10^3$ to $N_{\rm{q}}\simeq10^5$ in Fig.~\ref{num_vs_orac_vs_error} are fitted by $\log\hat{\epsilon}=\gamma\cdot\log N_{\rm{q}}+\delta$, and the fitted parameters corresponding to the slope are obtained as $\gamma=-0.76$, $\gamma=-0.95$ and $\gamma=-0.50$ for LIS and EIS, and classical random sampling, respectively. Similar slopes are obtained with other target probabilities. The fitted values of $\gamma$ for LIS and EIS are consistent with the slopes obtained using the Fisher information, although $\gamma$ slightly deviated from the theoretical values. This slight deviation indicates that $\hat{a}$ is a biased estimate; in fact, this deviation decreases as $N_{\textrm{shot}}$ increases, which is consistent with the fact that, in general, the ML estimate becomes unbiased asymptotically as the sampling number increases. Also, the efficiency of the ML estimate can be observed in the numerical simulation; the estimation error approaches the Cram\'er--Rao lower bound \eqref{lowest}. In Appendix A, we show the comparison of the error for the conventional phase-estimation-based approach with that of EIS. As a result, their estimation errors are found to be comparable.

\par
Finally, we remark that the computational complexity for naively finding the maximum of the likelihood function is on the order of $\mathcal{O}((1/\epsilon)\ln(1/\epsilon))$ if $m_k$ exponentially grows, as in EIS. This is because the computational complexity to obtain the likelihood function $\ln L(\vec{h};\theta_a)$ is evaluated as $\mathcal{O}(M)$ in this case. The order of the error $\epsilon$ is estimated as $\mathcal{O}(N_{\rm{q}}^{-1})$ based on the Cram\'er--Rao bound. Because $N_{\rm{q}}\sim 2^M N_{\rm{shot}}$, the complexity of evaluating the likelihood function is $\mathcal{O}(\ln (1/\epsilon))$. Assuming that the brute-force search among $1/\epsilon$ segments is performed to find the global maximum of the likelihood function, the complexity of finding the maximum is $\mathcal{{O}}((1/\epsilon)\ln(1/\epsilon))$. In the case of LIS, the order of the computational complexity can also be evaluated as $\mathcal{O}(\epsilon^{-5/3})$ in the same manner as before. It should be noted that the brute-force search algorithm for finding global minima of $\prod_{k=0}^{M}L_k(h_k;\theta_a)$ is not necessary if $m_k$ is zero for all $k$ (classical case), since the target value is simply obtained by $\hat{a}=\sum_{k=0}^{M}h_k/\sum_{k=0}^{M}N_{\rm{shot}}$. The error can be obtained as $\mathcal{O}(N_{\rm{q}}^{-1/2})$ based on the Cram\'{e}r--Rao  bound. Due to the fact that $N_{\rm{q}}=N_{\rm{shot}}M$, the computational complexity in the classical case is  $\mathcal{O}(\epsilon^{-2})$. The evaluated computational complexities of post-processing for different update rules of $m_k$ are summarized together with the query complexities in Table \ref{complexity}. 

\begin{table}[htb]
\begin{center}
  \caption{The summary of the complexities for estimating target value with given error $\epsilon$. The query complexity and computational complexity of post-processing for different update rules of $m_k$ are listed.}
\begin{tabular}{ccc}\hline
  update rule of $m_k$ & query complexity& computational complexity of \\
 & & post-processing\\ \hline
Classical \\
($m_k=0$ $\forall k$)& $\mathcal{O}(\epsilon^{-2})$ & $\mathcal{O}(\epsilon^{-2})$ \\ 
Linearly incremental sequence (LIS)\\
($m_0=0,m_1=1,m_2=2,\cdots,m_M=M$) & $\mathcal{O}(\epsilon^{-4/3})$ & $\mathcal{O}(\epsilon^{-5/3})$\\ 
Exponentially incremental sequence (EIS)\\
($m_0=0,m_1=2^0,m_2=2^1,\cdots,m_M=2^{(M-1)})$ & $\mathcal{O}(\epsilon^{-1})$ & $\mathcal{O}(\epsilon^{-1}\ln \epsilon^{-1})$\\
\hline
\label{complexity}
\end{tabular}
\end{center}
\end{table}

\section{Application to the Monte Carlo integration}
We conduct a Monte Carlo integration as an example of the application of our algorithm, as follows. In this section, we first review the quantum algorithm to calculate the Monte Carlo integration by amplitude estimation \cite{Montanaro2015} and then explain the amplitude amplification operator used in our algorithm. Next, we present the integral of the sine function as a simple example of Monte Carlo integration. Using this example, we discuss the number of CNOT gates and qubits required for our algorithm and the conventional amplitude estimation \cite{Brassard2002}.

\subsection{The Monte Carlo integration as an amplitude estimation}
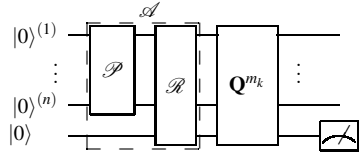
\begin{figure}[tb]
		\centering
		\leavevmode
		\Qcircuit @C=1em @R=.3em {
		& & \hspace{10mm}\mathcal{A} & \nghost{} & & & \\
 & \lstick{\ket{0}^{(1)}} & \multigate{5}{\mathcal P} & \multigate{6}{\mathcal R} & \multigate{6}{\mathbf{Q}^{m_k}} & \qw & \qw \\
		\\
		& \lstick{\vdots} & \nghost{\mathcal P} & \nghost{\mathcal R} & & \vdots \\
		\\
		\\
 & \lstick{\ket{0}^{(n)}} & \ghost{\mathcal P} & \ghost{\mathcal R} & \ghost{\mathcal{Q}^{m_k}} & \qw & \qw \\
		& \lstick{\ket{0}_{\hspace{1.15em} }} & \qw & \ghost{\mathcal R} & \ghost{\mathcal{Q}^{m_k}} & \qw & \meter
		\gategroup{2}{3}{8}{4} {.3em}{--}
		}
\caption{Quantum circuit of amplitude amplification for the Monte Carlo integration.}
		\label{fig:circuit_Grover_schematic}
	\end{figure}

\begin{figure}[tb]
		\centering
		\leavevmode
		\Qcircuit @C=1em @R=.3em {
 & \lstick{\ket{0}^{(1)}} & \multigate{5}{\mathcal P} & \multigate{6}{\mathcal R} & \multigate{6}{\mathbf Q} & \multigate{6}{{\mathbf Q}^{2}} & \qw & \ldots & & \multigate{6}{{\mathbf Q}^{2^{m-1}}} & \qw & \qw \\
		\\
		& \lstick{\vdots} & \nghost{\mathcal P} & \nghost{\mathcal R} & & & & \cdots & & & \vdots \\
		\\
		\\
 & \lstick{\ket{0}^{(n)}} & \ghost{\mathcal P} & \ghost{\mathcal R} & \ghost{\mathbf{Q}} & \ghost{\mathbf{Q}^{2}} & \qw & \ldots & & \ghost{\mathbf{Q}^{2^{m-1}}} & \qw & \qw \\
 & \lstick{\ket{0}_{\hspace{1.15em} }} & \qw & \ghost{\mathcal R} & \ghost{\mathbf{Q}} & \ghost{\mathbf{Q}^{2}} & \qw & \ldots & & \ghost{\mathbf{Q}^{2^{m-1}}} & \qw & \qw & \\
		\\
		\\
 & \lstick{\ket{0}^{(1)}} & \gate{H} & \qw & \ctrl{-3} & \qw & \qw & \ldots & & \qw & \multigate{6}{F_{m}^{-1}} & \meter \\
 & \lstick{\ket{0}^{(2)}} & \gate{H} & \qw & \qw & \ctrl{-4} & \qw & \ldots & & \qw & \ghost{F_{m}^{-1}} & \meter \\
		\\
		& \lstick{\vdots} & & & & & & \ddots & & & & \vdots \\
		\\
		\\
 & \lstick{\ket{0}^{(m)}} & \gate{H} & \qw & \qw & \qw & \qw & \ldots & & \ctrl{-9} & \ghost{F_{m}^{-1}} & \meter
		\\
		}
\caption{Quantum circuit of conventional amplitude estimation for the Monte Carlo integration. $F_{m}^{-1}$ represents the inverse QFT of $m$ qubits.}
		\label{fig:circuit_amplitudeEstimation_schematic}
	\end{figure}
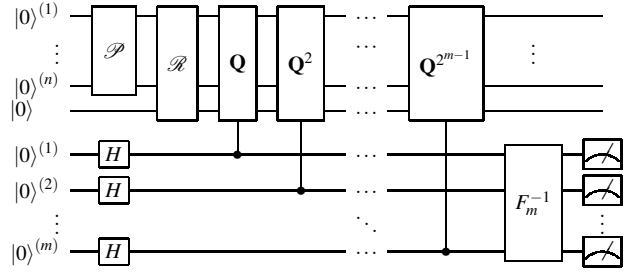

One purpose of the Monte Carlo integration is to calculate the expected value of real valued function $ 0 \le f(x) \le 1$ defined for $n$-bit input $x \in \{0,1\}^n$ with probability $p(x)$:
	\begin{equation}
		\mathbb{E}[f(x)]=\sum_{x=0}^{2^n-1}p(x)f(x). \label{expectation_value}
	\end{equation}
In the quantum algorithm for the Monte Carlo integration, an additional (ancilla) qubit is introduced and assumed to be rotated as
	\begin{equation}
		\mathcal{R}\ket{x}_n\ket{0}=\ket{x}_n\left(\sqrt{f(x)}\ket{1}+\sqrt{1-f(x)}\ket{0}\right),\label{rotation}
	\end{equation}
where $\mathcal{R}$ is a unitary operator acting on $n+1$ qubits. In addition, an algorithm $\mathcal{P}$ is introduced, and operating $\mathcal{P}$ to $n$-qubit resister $\ket{0}_n$ yields
	\begin{equation}
		\mathcal{P}\ket{0}_n=\sum_{x=0}^{2^n-1}\sqrt{p(x)}\ket{x}_n,
	\end{equation}
where all qubits in $\ket{0}_n$ are in the state $\ket{0}$. Operating $\mathcal{R(P\otimes\mathbf{I}_{\rm{1}})}$ to the state $\ket{0}_n\ket{0}$ generates $\ket{\Psi}$:
	\begin{eqnarray}
		\ket{\Psi}&=&\mathcal{R(P\otimes\mathbf{I}_{\rm{1}})}\ket{0}_n\ket{0}\\
		&=&\sum_{x=0}^{2^n-1}\sqrt{p(x)}\ket{x}_n\left(\sqrt{f(x)}\ket{1}+\sqrt{1-f(x)}\ket{0}\right),\label{chi_vec}
	\end{eqnarray}
where $\mathbf{I}_{\rm{1}}$ is the identity operator acting on an ancilla qubit. For convenience, we put $a=\sum_{x=0}^{2^n-1}p(x)f(x)$ and introduce two orthonormal bases:
	\begin{eqnarray}
		\ket{\tilde{\Psi}_1}&=&\frac{1}{\sqrt{a}}\sum_{x=0}^{2^n-1}\sqrt{p(x)}\sqrt{f(x)}\ket{x}_n\ket{1},\label{chi1}\\
		\ket{\tilde{\Psi}_0}&=&\frac{1}{\sqrt{1-a}}\sum_{x=0}^{2^n-1}\sqrt{p(x)}\sqrt{1-f(x)}\ket{x}_n\ket{0}.\label{chi0}
	\end{eqnarray}
By using these bases, the state $\ket{\Psi}$ can be rewritten as
	\begin{equation}
		\ket{\Psi}=\sqrt{a} \ket{\tilde{\Psi}_1}+\sqrt{1-a} \ket{\tilde{\Psi}_0}.\label{chi_chi0_chi1}
	\end{equation}
  Then, the square root of expected value $a=\mathbb{E}[f(x)]$ appears in the amplitude of $\ket{\tilde{\Psi}_1}$, and the Monte Carlo integration can be regarded as an amplitude estimation of $\ket{\tilde{\Psi}_1}$. The operator $\mathbf{Q}$ defined in Eq.~(\ref{Q_def}) can be achieved using $\mathbf{U}_{\Psi}\mathbf{U}_{\tilde{\Psi}_0}$, where $\mathbf{U}_{\tilde{\Psi}_0}=\mathbf{I}-2\ket{\tilde{\Psi}_0}\bra{\tilde{\Psi}_0}$, $\mathbf{U}_{\Psi}=\mathbf{I}-2\ket{\Psi}\bra{\Psi}$, and $\mathbf{I}$ is the identity acting on $n+1$ qubits \cite{Brassard2002}. In terms of a practical point of view, we use $\mathbf{U}_{\tilde{\Psi}_0}=\mathbf{I}_{n+1}-2\mathbf{I}_n\ket{0}\bra{0}$, where $\mathbf{I}=\mathbf{I}_{n+1}=\mathbf{I}_n\otimes(\ket{0}\bra{0}+\ket{1}\bra{1})$. By putting $a=\sin^2\theta_a$ and using Eq.~(\ref{Qj2}), we could apply our algorithm to the Monte Carlo integration. The circuit diagram of the amplitude amplification used in our algorithm is shown in Fig.~\ref{fig:circuit_Grover_schematic}. Note that the multi-qubit gate consisting of $\mathcal{P}$ and $\mathcal{R}$ in Fig.~\ref{fig:circuit_Grover_schematic} corresponds to the quantum algorithm $\mathcal{A}$ shown in Sec.~\ref{sec:preliminary}, and the only ancilla qubit for each $k$ is measured when our algorithm is applied to the Monte Carlo. Similarly, the circuit of the conventional amplitude estimation \cite{Brassard2002} is shown in Fig.~\ref{fig:circuit_amplitudeEstimation_schematic}. In the following, we applied our algorithm to a very simple integral of the sine function and compared the number of CNOT gates and qubits with the results of the conventional amplitude estimation.
\subsection{Simple example: integral of the sine function}
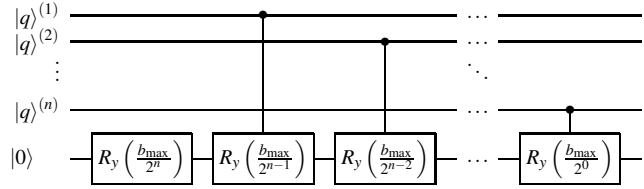
\begin{figure}[tb]
		\centering
		\small
		\leavevmode
		\Qcircuit @C=1em @R=1em {
 \lstick{\ket{q}^{(1)}} & \qw & \ctrl{5} & \qw & \qw & \ldots & &\qw & \qw \\
 \lstick{\ket{q}^{(2)}} & \qw & \qw & \ctrl{4} & \qw & \ldots & &\qw & \qw \\
 \lstick{\vdots} & & & & & \ddots\\
 \\
 \lstick{\ket{q}^{(n)}} & \qw & \qw & \qw &\qw & \ldots & & \ctrl{1} & \qw \\
 \lstick{\ket{0}_{\hspace{1.15em}}} & \gate{R_y\left(\frac{b_{\textrm{max}}}{2^n}\right)} & \gate{R_y\left(\frac{b_{\textrm{max}}}{2^{n-1}}\right)} & \gate{R_y\left(\frac{b_{\textrm{max}}}{2^{n-2}}\right)} &\qw & \ldots & & \gate{R_y\left(\frac{b_{\textrm{max}}}{2^0}\right)} &\qw
		}
\caption{Quantum circuit achieving the operator $\mathcal{R}$ in Eq.~\eqref{eq:Rotation_SimpleSin}. In this circuit, $\ket{x}$ in Eq.~\eqref{eq:Rotation_SimpleSin} is represented by $n$ qubits, denoted by $\ket{q}^{(1)}$, $\ket{q}^{(2)}$,$\cdots$, $\ket{q}^{(n)}$. $R_y(\theta)$ represents a Y-rotation with angle $\theta$.}
		\label{fig:circuit_R}
	\end{figure}

As a simple example of the Monte Carlo integration, the following integral is considered.
	\begin{equation}
 I = \frac{1}{b_{\textrm{max}}} \int_0^{b_{\textrm{max}}} \sin(x)^2 dx,
	\end{equation}
where $b_{\textrm{max}}$ is a constant that determines the upper limit of the integral. By discretizing this integral in $n$-qubit, we obtain
	\begin{equation}
 S = \sum_{x=0}^{2^n-1} p\left(x\right) \sin^2\left(\frac{\left(x+\frac{1}{2}\right)b_{\textrm{max}}}{2^n}\right), \label{eq:SumSimpleSin}
	\end{equation}
where $p(x)=\frac{1}{2^n}$ is a discrete uniform probability distribution. We now explicitly describe the operators $\mathcal{P}$ and $\mathcal{R}$ for applying our algorithm to calculate the sum \eqref{eq:SumSimpleSin}. The operator $\mathcal{P}$ acting on the $n$-qubit initial state can be defined as
	\begin{equation}
		\mathcal{P}: \ket{0}_n \ket{0} \to \frac{1}{\sqrt{2^n}} \sum_{x=0}^{2^n-1}\ket{x}_n\ket{0}. \label{eq:A_SimpleSin}
	\end{equation}
The operator $\mathcal{P}$ can be constructed using $n$ Hadamard gates. The operator $\mathcal{R}$ acting on the $(n+1)$-qubit state $\ket{x}_n\ket{0}$ can be defined as
	\begin{equation}
		\mathcal{R}: \ket{x}_n \ket{0} \to \ket{x}_n\left(		
 \sin\left( \frac{\left(x+\frac{1}{2}\right)b_{\textrm{max}}}{2^n} \right)\ket{1}
 +\cos\left( \frac{\left(x+\frac{1}{2}\right)b_{\textrm{max}}}{2^n} \right)\ket{0}
 \right). \label{eq:Rotation_SimpleSin}
	\end{equation}
The operator $\mathcal{R}$ can be constructed using controlled Y-rotations as illustrated in Fig.~\ref{fig:circuit_R}.

We now explicitly show an example of the circuits for the amplitude amplification used in our algorithm and a conventional amplitude estimation with a single $\mathbf{Q}$ operation, which calculates the sum \eqref{eq:SumSimpleSin}. For simplicity, the circuit for $b_{\textrm{max}} = \pi/4 $ and $n=2$ is shown here. The quantum circuits for amplitude amplification and conventional amplitude estimation are shown in Fig.~\ref{fig:circuit_Grover} and Fig.~\ref{fig:circuit_amplitudeEstimation}, respectively. In these circuits, all-to-all qubit connectivity is assumed. From these figures, we can see that the circuit for conventional amplitude estimation tends to have more gates and qubits than that of our algorithm. Furthermore, the multi-controlled operation in the conventional amplitude estimation circuit of Fig.~\ref{fig:circuit_amplitudeEstimation} may require several ancilla qubits. 

Table~\ref{tab:num_of_CX_QUBIT} shows the number of CNOT gates and qubits as a function of the number of $\mathbf{Q}$ operators required for conventional amplitude estimation and our algorithm. Here, we assume the gate set supported by Qiskit ver. 0.7 \cite{Qiskit}. Because the number of CNOT gate operations is restricted in NISQ devices due to the error accumulation, the numbers of CNOT gates in our algorithm only those for the circuit with the largest $m_k$ are evaluated. The numbers of CNOT gates in our algorithm are about 7--18 times smaller than those of conventional amplitude estimation. The number of qubits required for conventional amplitude estimation increases as the number of $\mathbf{Q}$ operations increased, while that for our algorithm kept constant. The source code for Monte Carlo integration based on our proposed algorithm is available at \cite{Github}.

	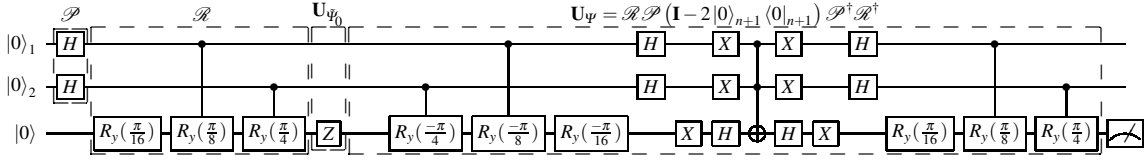
\begin{figure}[tb]
		\centering
		\resizebox{.99\linewidth}{!}{
		\leavevmode
		\scriptsize
		\Qcircuit @C=.5em @R=.8em {
 & \mathcal{P} & & \mathcal{R} & & \mathbf{U}_{\tilde{\Psi}_0} & & & & & & & \mathbf{U}_{\Psi} = \mathcal{R} \mathcal{P}\left( \mathbf{I} -2\ket{0}_{n+1}\bra{0}_{n+1} \right) \mathcal{P}^\dagger \mathcal{R}^\dagger \\
	\lstick{\ket{0}_1} & \gate{H} & \ggate{} & \ctrl{2} & \qw & \ggate{} & \ggate{} & \qw & \ctrl{2} & \qw & \gate{H} & \ggate{} & \gate{X} & \ctrl{1} & \gate{X} & \ggate{} & \gate{H} & \qw & \ctrl{2} & \qw & \qw\\
	\lstick{\ket{0}_2} & \gate{H} & \ggate{} & \ggate{} & \ctrl{1} & \ggate{} & \ggate{} & \ctrl{1} & \ggate{} & \ggate{} & \gate{H} & \ggate{} & \gate{X} & \ctrl{1} & \gate{X} & \ggate{} & \gate{H} & \ggate{} & \ggate{} & \ctrl{1} & \qw\\
	\lstick{\ket{0}} & \ggate{} & \gate{R_y(\frac{\pi}{16})} & \gate{R_y(\frac{\pi}{8})} & \gate{R_y(\frac{\pi}{4})} & \gate{Z} & \ggate{} & \gate{R_y(\frac{-\pi}{4})} & \gate{R_y(\frac{-\pi}{8})} & \gate{R_y(\frac{-\pi}{16})} & \ggate{} & \gate{X} & \gate{H} & \targ & \gate{H} & \gate{X} & \ggate{} & \gate{R_y(\frac{\pi}{16})} & \gate{R_y(\frac{\pi}{8})} & \gate{R_y(\frac{\pi}{4})} & \meter
		\gategroup{2}{2}{3}{2} {.3em}{--}
		\gategroup{2}{3}{4}{5} {.3em}{--}
		\gategroup{2}{6}{4}{6} {.3em}{--}
		\gategroup{2}{7}{4}{20}{.3em}{--}
		}
		}
\caption{Quantum circuit of amplitude amplification in the case of $n=2$ with single $\mathbf{Q}$ operation.}
		\label{fig:circuit_Grover}
	\end{figure}
	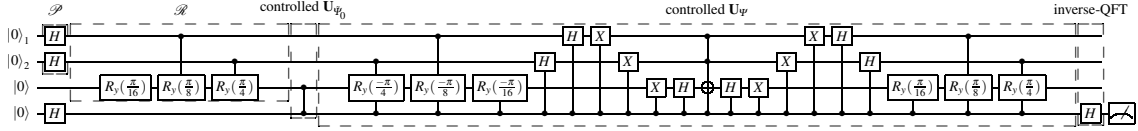
\begin{figure}[tb]
		\centering
		\resizebox{.99\linewidth}{!}{
		\leavevmode
		\scriptsize
		\Qcircuit @C=0.5em @R=0.0em @!R {
		& \mathcal{P} & & & \mathcal{R} & & & \textrm{controlled } \mathbf{U}_{\tilde{\Psi}_0} & & & & & & & & & & & \textrm{controlled } \mathbf{U}_{\Psi} & & & & & & & & & & &\textrm{inverse-QFT} \\
    \lstick{\ket{0}_1} & \gate{H} & \ggate{} & \qw & \ctrl{2} & \qw & \qw & \ggate{} & \ggate{} & \qw & \ctrl{2} & \qw & \ggate{} & \gate{H} & \gate{X} & \ggate{} & \ggate{} & \qw & \ctrl{1} & \qw & \ggate{} & \ggate{} & \gate{X} & \gate{H} & \ggate{} & \qw & \ctrl{2} & \qw & \qw & \ggate{}\\
	\lstick{\ket{0}_2} & \gate{H} & \ggate{} & \ggate{} & \ggate{} & \ctrl{1} & \qw & \ggate{} & \ggate{} & \ctrl{1} & \ggate{} & \ggate{} & \gate{H} & \ggate{} & \ggate{} & \gate{X} & \ggate{} & \qw & \ctrl{1} & \qw & \ggate{} & \gate{X} & \ggate{} & \ggate{} & \gate{H} & \ggate{} & \ggate{} & \ctrl{1} & \qw & \ggate{} \\
	\lstick{\ket{0}} & \ggate{} & \ggate{} & \gate{R_y(\frac{\pi}{16})} & \gate{R_y(\frac{\pi}{8})} & \gate{R_y(\frac{\pi}{4})} & \ggate{} & \control\qw & \qw & \gate{R_y(\frac{-\pi}{4})} & \gate{R_y(\frac{-\pi}{8})} & \gate{R_y(\frac{-\pi}{16})} & \ggate{} & \ggate{} & \qw & \ggate{} & \gate{X} & \gate{H} & \targ & \gate{H} & \gate{X} & \ggate{} & \qw & \ggate{} & \ggate{} & \gate{R_y(\frac{\pi}{16})} & \gate{R_y(\frac{\pi}{8})} & \gate{R_y(\frac{\pi}{4})} & \qw & \ggate{} \\
	\lstick{\ket{0}} & \gate{H} & \qw & \qw & \qw & \qw & \ggate{} & \ctrl{-1} & \qw & \ctrl{-1} & \ctrl{-1} & \ctrl{-1} & \ctrl{-2} & \ctrl{-3} & \ctrl{-3} & \ctrl{-2} & \ctrl{-1}& \ctrl{-1} & \ctrl{-3} & \ctrl{-1} & \ctrl{-1}& \ctrl{-2} & \ctrl{-3} & \ctrl{-3} & \ctrl{-2} & \ctrl{-1} & \ctrl{-1} & \ctrl{-1} & \ggate{} & \gate{H} & \meter
		\gategroup{2}{2}{3}{2} {.3em}{--}
		\gategroup{2}{3}{4}{7} {.3em}{--}
		\gategroup{2}{8}{5}{8} {.3em}{--}
		\gategroup{2}{9}{5}{29}{.3em}{--}
		\gategroup{2}{30}{5}{30}{.3em}{--}
		}
		}
\caption{Quantum circuit of conventional amplitude estimation in the case of $n=2$ with a single $\mathbf{Q}$ operation.}
		\label{fig:circuit_amplitudeEstimation}
	\end{figure}\par
\begin{table}[tb]
	\begin{center}
\caption{Number of CNOT gates and qubits to calculate \eqref{eq:SumSimpleSin} as a function of $\mathbf{Q}$ operations.}
		\begin{tabular}{rrrrr} \hline
			 & \multicolumn{2}{c}{conventional amplitude estimation} & \multicolumn{2}{c}{our algorithm} \\
			\# operators $\mathbf{Q}$ & \# CNOT gates & \# qubits & \# CNOT gates & \# qubits \\\hline
			0 & - & - & 4 & 3 \\
			$2^0$ & 135 & 7 & 18 & 3 \\
			$2^1$ & 399 & 8 & 32 & 3 \\
			$2^2$ & 927 & 9 & 60 & 3 \\
			$2^3$ & 1981 & 10 & 116 & 3 \\
			$2^4$ & 4085 & 11 & 228 & 3 \\
			$2^5$ & 8287 & 12 & 452 & 3 \\
			$2^6$ & 16683 & 13 & 900 & 3 \\
			$2^7$ & 33465 & 14 & 1796 & 3 \\
			$2^8$ & 67017 & 15 & 3588 & 3 \\\hline
		\end{tabular}
		\label{tab:num_of_CX_QUBIT}
	\end{center}
\end{table}

\section{Conclusion}
We proposed a quantum amplitude estimation algorithm achieving quantum speedup by reducing controlled gates with ML estimation. The essential idea of the proposed algorithm is constructing a likelihood function using the outcomes of measurements on several quantum states, which are transformed by the amplitude amplification process. Although the probability measuring good or bad states depends on the number of amplitude amplification operations, the outcomes are correlated due to the fact that each amplified probability is a function of a single parameter. To test the efficiency of the proposed algorithm, we performed numerical simulations, and analyzed the relationships between the number of queries and estimation error. Empirical evidences showed the algorithm could estimate the target value with fewer queries than the classical algorithm. We also presented the lower bound of the estimation error in terms of the Fisher information and found that the estimation error observed in a numerical simulation was sufficiently close to the Heisenberg limit. In addition, we experimented the proposed algorithm for a Monte Carlo integration, and found that fewer CNOT gates and qubits were required in comparison with the conventional amplitude estimation. These facts indicate that our algorithm could work well even with noisy intermediate-scale quantum devices. 

Shortly after the publication of our results, simplified quantum counting and amplitude estimation without QFT with rigorous proofs were shown \cite{Aaronson2019}. In contrast to our approach that can be run in parallel on multiple quantum devices, the simplified algorithms are adaptive and have to be run sequentially. They also require large constant-factor overhead, e.g., millions of measurement samples, which could be expensive in practice. Nevertheless, there are several interesting directions for future work as pointed out in \cite{Aaronson2019}, such as, obtaining rigorous proofs for our parallel approach and achieving quantum speedups with depth-limited quantum circuits.

\begin{acknowledgements}
We thank Yutaka Shikano and Hideo Watanabe for their constructive comments. This work was supported by MEXT Quantum Leap Flagship Program Grant Number JPMXS0118067285.
\end{acknowledgements}

\begin{appendices}
\renewcommand{\thefigure}{\Alph{section}}
\section{Comparison of estimation errors with conventional amplitude estimation}
We compare the estimation error between the conventional amplitude estimation algorithm and our proposed algorithm.  The details of conventional amplitude estimation algorithm is presented in Ref.~\cite{Brassard2002}. For simplicity, only the result of $a = \sin^2{\theta_a} = 1/48$ is shown here.

\begin{figure}[htbp]
  \centering
  \includegraphics{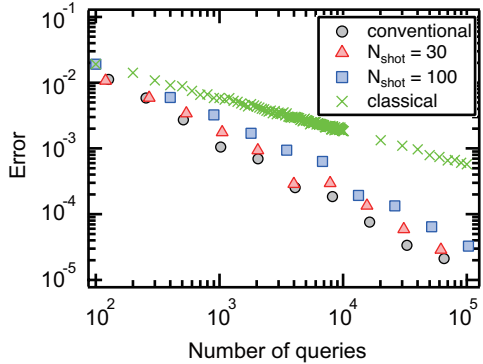}
  \caption{The relationship between the number of queries and estimation error of our proposed algorithm and conventional amplitude estimation~\cite{Brassard2002}. The (black) circles are from the conventional phase-estimation-based approach. The (red) triangles and (blue) squares are the 81 percentile values of estimation error with numerical simulations for $N_{\rm{shot}}=30$ and $N_{\rm{shot}}=100$ for EIS, respectively. For comparison, the 81 percentile values of estimation error with classical sampling are also shown as (green) crosses. Against a fixed total number of queries, the smaller the $N_{\rm{shot}}$, the more queries are used for the quantum amplitude amplification, and hence more speedup approaching the conventional phase-estimation.}
  \label{fig:conventional_error}
\end{figure}

Fig.~\ref{fig:conventional_error} shows relations between the number of queries and estimation error between the conventional amplitude estimation and our proposed algorithm.
In the figure, the (black) circles, which represent the data of conventional algorithm, are generated in the following manner.
The conventional algorithm outputs four integers closest to the target value $\theta_a M/\pi$ and $M - \theta_a M/\pi$ with success probability of at least $8/\pi^2 \times 100 \% \sim 81\%$ after $M=2^m-1$ times application of controlled $\mathbf{Q}$ operation followed by QFT~\cite{Brassard2002}.
The largest of the estimation error calculated from these four integers is plotted in Fig.~\ref{fig:conventional_error}.
The (red) triangles and (blue) squares represent the data of our proposed method with $N_{\textrm{shot}}=30$ and $100$, respectively.
The $8/\pi^2 \times 100 \sim 81$ percentile of the estimation error is plotted here for a fair comparison with the conventional algorithm, although the averaged error is described in the main text. The data of our algorithm is generated by the same manner as in Section 3.3.

This figure shows that the estimation error of our proposed method increases gradually as the number of shots increases. 
This is because, as can be seen from Eqs.~\ref{Fisher_final} and \ref{calling}, the degree of quantum speedup becomes relatively smaller by increasing the number of shots to the limit that it is essentially a classical sampling when the number of shots is equal to the total number of queries. The figure shows the estimation error of the conventional algorithm is almost the same as that of ours with the small $N_{\textrm{shot}}=30$.

\end{appendices}

\newpage
%\bibliographystyle{spmpsci_unsort}
%\bibliography{reference}

\begin{thebibliography}{10}
\providecommand{\url}[1]{{#1}}
\providecommand{\urlprefix}{URL }
\expandafter\ifx\csname urlstyle\endcsname\relax
  \providecommand{\doi}[1]{DOI~\discretionary{}{}{}#1}\else
  \providecommand{\doi}{DOI~\discretionary{}{}{}\begingroup
  \urlstyle{rm}\Url}\fi

\bibitem{IBMQ2019}
{IBM Q Experience}.
\newblock \url{https://quantumexperience.ng.bluemix.net/qx/editor}.
\newblock Accessed: 2019-03-26

\bibitem{PhysRevX.8.021012}
Friis, N., Marty, O., Maier, C., Hempel, C., Holz\"apfel, M., Jurcevic, P.,
  Plenio, M.B., Huber, M., Roos, C., Blatt, R., Lanyon, B.: Observation of
  entangled states of a fully controlled 20-qubit system.
\newblock Phys. Rev. X \textbf{8}, 021012 (2018)

\bibitem{PhysRevLett.119.180511}
Song, C., Xu, K., Liu, W., Yang, C.p., Zheng, S.B., Deng, H., Xie, Q., Huang,
  K., Guo, Q., Zhang, L., Zhang, P., Xu, D., Zheng, D., Zhu, X., Wang, H.,
  Chen, Y.A., Lu, C.Y., Han, S., Pan, J.W.: 10-qubit entanglement and parallel
  logic operations with a superconducting circuit.
\newblock Phys. Rev. Lett. \textbf{119}, 180511 (2017)

\bibitem{Preskill2018}
Preskill, J.: Quantum {C}omputing in the {NISQ} era and beyond.
\newblock {Quantum} \textbf{2}, 79 (2018)

\bibitem{McClean2016}
McClean, J.R., Romero, J., Babbush, R., Aspuru-Guzik, A.: The theory of
  variational hybrid quantum-classical algorithms.
\newblock New J. Phys. \textbf{18}, 023023 (2016)

\bibitem{Yung2014}
Yung, M.H., Casanova, J., Mezzacapo, A., McClean, J., Lamata, L., Aspuru-Guzik,
  A., Solano, E.: From transistor to trapped-ion computers for quantum
  chemistry.
\newblock Sci. Rep. \textbf{4}, 3589 (2014)

\bibitem{Knill2007}
Knill, E., Ortiz, G., Somma, R.D.: Optimal quantum measurements of expectation
  values of observables.
\newblock Phys. Rev. A \textbf{75}, 012328 (2007)

\bibitem{Guzik2008}
Kassala, I., Jordan, S.P., Lovec, P.J., Mohsenia, M., Aspuru-Guzik, A.:
  Polynomial-time quantum algorithm for the simulation of chemical dynamics.
\newblock Proc. Natl. Acad. Sci. USA \textbf{105}, 18681--18686 (2008)

\bibitem{Bromley2018}
Rebentrost, P., Gupt, B., Bromley, T.R.: Quantum computational finance: {Monte
  Carlo} pricing of financial derivatives.
\newblock Phys. Rev. A \textbf{98}, 022321 (2018)

\bibitem{Woerner2019}
Woerner, S., Egger, D.J.: Quantum risk analysis.
\newblock npj Quantum Inf. \textbf{5}, 15 (2019)

\bibitem{Wiebe2015}
Wiebe, N., Kapoor, A., Svore, K.M.: Quantum algorithms for nearest-neighbor
  methods for supervised and unsupervised learning.
\newblock Quantum Inf. Comput. \textbf{15}, 316--356 (2015)

\bibitem{Wiebe2016a}
Wiebe, N., Kapoor, A., Svore, K.M.: Quantum deep learning.
\newblock Quantum Inf. Comput. \textbf{16}, 541--587 (2016)

\bibitem{Wiebe2016b}
Wiebe, N., Kapoor, A., Svore, K.M.: Quantum perceptron models.
\newblock Proceedings of the 30th International Conference on Neural
  Information Processing Systems pp. 4006--4014 (2016)

\bibitem{Kerenidis2018}
Kerenidis, I., Landman, J., Luongo, A., Prakash, A.: q-means: A quantum
  algorithm for unsupervised machine learning.
\newblock arXiv:1812.03584  (2018)

\bibitem{Montanaro2015}
Montanaro, A.: Quantum speedup of {Monte Carlo} methods.
\newblock Proc. Royal Soc. A \textbf{471}, 20150301 (2015)

\bibitem{Brassard2002}
Brassard, G., H{\o}yer, P., Mosca, M., Tapp, A.: Quantum amplitude
  amplification and estimation.
\newblock Contemporary Mathematics Series Millenium \textbf{305}, 53--74 (2002)

\bibitem{kitaev1996quantum}
Kitaev, A.Y.: Quantum measurements and the {Abelian} stabilizer problem.
\newblock Electronic Colloquium on Computational Complexity \textbf{3} (1996)

\bibitem{Svore2013}
Svore, K.M., Hastings, M.B., Freedman, M.: Faster phase estimation.
\newblock Quantum Inf. Comput. \textbf{14}, 306--328 (2014)

\bibitem{Granade2016}
Wiebe, N., Granade, C.: Efficient {Bayesian} phase estimation.
\newblock Phys. Rev. Lett. \textbf{117}, 010503 (2016)

\bibitem{OBrien2018}
O'Brien, T.E., Tarasinski, B., Terhal, B.M.: Quantum phase estimation of
  multiple eigenvalues for small-scale (noisy) experiments.
\newblock New J. Phys. \textbf{21}, 023022 (2019)

\bibitem{denBerg2019}
van~den Berg, E.: Practical sampling schemes for quantum phase estimation.
\newblock arXiv:1902.11168  (2019)

\bibitem{CRWie2019}
Wie, C.R.: Simpler Quantum Counting 
\newblock arXiv:1907.08119  (2019)

\bibitem{Grover1996}
Grover, L.K.: A fast quantum mechanical algorithm for database search.
\newblock Proceedings of 28th Annual ACM Symposium on Theory of Computing pp.
  212--219 (1996)

\bibitem{Abrams1999}
Abrams, D.S., Williams, C.P.: Fast quantum algorithms for numerical integrals
  and stochastic processes.
\newblock arXiv:quant-ph/9908083  (1999)

\bibitem{Zintchenko2016}
Zintchenko, I.,Wiebe, N.: Randomized gap and amplitude estimation.
\newblock Phys. Rev. A \textbf{93}, 062306 (2016)

\bibitem{Brassard1997}
Brassard, G., H{\o}yer, P.: An exact quantum polynomial-time algorithm for
  {Simon's} problem.
\newblock Proceedings of the 5th Israeli Symposium on Theory of Computing and
  Systems pp. 12--23 (1997)

\bibitem{Grover1998}
Grover, L.K.: Quantum computers can search rapidly by using almost any
  transformation.
\newblock Phys. Rev. Lett. \textbf{80}, 4329--4332 (1998)

%\bibitem{Brassard2000}
%Brassard, G., H{\o}yer, P., Mosca, M., Tapp, A.: Quantum amplitude
%  amplification and estimation.
%\newblock Contemporary Mathematics \textbf{305}, 53--74 (2002)

\bibitem{rao1973linear}
Rao, C.R.: Linear statistical inference and its applications, vol.~2.
\newblock Wiley New York (1973)

\bibitem{Qiskit}
Aleksandrowicz, G., Alexander, T., Barkoutsos, P., Bello, L., Ben-Haim, Y.,
  Bucher, D., Cabrera-Hern{\'a}dez, F.J., Carballo-Franquis, J., Chen, A.,
  Chen, C.F., Chow, J.M., C{\'o}rcoles-Gonzales, A.D., Cross, A.J., Cross, A.,
  Cruz-Benito, J., Culver, C., Gonz{\'a}lez, S.D.L.P., Torre, E.D.L., Ding, D.,
  Dumitrescu, E., Duran, I., Eendebak, P., Everitt, M., Sertage, I.F., Frisch,
  A., Fuhrer, A., Gambetta, J., Gago, B.G., Gomez-Mosquera, J., Greenberg, D.,
  Hamamura, I., Havlicek, V., Hellmers, J., Herok, {\L}., Horii, H., Hu, S.,
  Imamichi, T., Itoko, T., Javadi-Abhari, A., Kanazawa, N., Karazeev, A.,
  Krsulich, K., Liu, P., Luh, Y., Maeng, Y., Marques, M.,
  Mart{\'\i}n-Fern{\'a}ndez, F.J., McClure, D.T., McKay, D., Meesala, S.,
  Mezzacapo, A., Moll, N., Rodr{\'\i}guez, D.M., Nannicini, G., Nation, P.,
  Ollitrault, P., O'Riordan, L.J., Paik, H., P{\'e}rez, J., Phan, A., Pistoia,
  M., Prutyanov, V., Reuter, M., Rice, J., Davila, A.R., Rudy, R.H.P., Ryu, M.,
  Sathaye, N., Schnabel, C., Schoute, E., Setia, K., Shi, Y., Silva, A.,
  Siraichi, Y., Sivarajah, S., Smolin, J.A., Soeken, M., Takahashi, H.,
  Tavernelli, I., Taylor, C., Taylour, P., Trabing, K., Treinish, M., Turner,
  W., Vogt-Lee, D., Vuillot, C., Wildstrom, J.A., Wilson, J., Winston, E.,
  Wood, C., Wood, S., W{\"o}rner, S., Akhalwaya, I.Y., Zoufal, C.: Qiskit: An
  open-source framework for quantum computing (2019).
\newblock \doi{10.5281/zenodo.2562110}


\bibitem{Github}
{Qiskit Community Tutorials: 
Amplitude Estimation without Quantum Fourier Transform and Controlled Grover Operators}.
\newblock \texttt{https:{\slash}{\slash}github.com{\slash}Qiskit{\slash}qiskit-community-tutorials{\slash}blob{\slash}master{\slash}algorithms{\slash}SimpleIntegral\_AEwoPE.ipynb}.
\newblock Accessed: 2019-10-30


\bibitem{Aaronson2019}
Aaronson, S., Rall, P.: Quantum Approximate Counting, Simplified 
\newblock arXiv:1908.10846  (2019)


\end{thebibliography}

\end{document}